\documentclass{article}
\usepackage{algorithm,algpseudocode,bm,cancel,import,spconf,amsmath,amssymb,graphicx,verbatim,url,xcolor,soul}

\newcommand\lpipe{\mkern1.5mu{|}\mkern1.5mu}
\newcommand\lstar{\mkern4mu{\star}\mkern4mu}

\title{Low SNR Multiframe Registration for Cubesats}
\name{Evan Widloski and Farzad Kamalabadi}
\address{University of Illinois Urbana-Champaign}

\begin{document}
\maketitle
\begin{abstract}
  We present a registration algorithm which jointly estimates motion and the ground truth image from a set of noisy frames under rigid, constant translation. The algorithm is non-iterative and needs no hyperparameter tuning.  It requires a fixed number of FFT, multiplication, and downsampling operations for a given input size, enabling fast implementation on embedded platforms like cubesats where on-board image fusion can greatly save on limited downlink bandwidth.  The algorithm is optimal in the maximum likelihood sense for additive white Gaussian noise and non-stationary Gaussian approximations of Poisson noise.  Accurate registration is achieved for very low SNR, even when visible features are below the noise floor.
\end{abstract}


\section{Introduction}
\label{sec:introduction}

Image alignment, or image registration is a classic problem in the field of image processing involving the alignment of an image set or image sequence that has been acquired at different times, sensors, or viewpoints.
Registration is an important preprocessing step in image processing pipelines involving change detection, image mosaicing, denoising, and super-resolution with applications in medical imaging \cite{wells1996multi}, computer vision tasks \cite{mers} like segmentation or classification, military surveillance \cite{konrad}, and remote-sensing.


Registration techniques have been broadly classified into \emph{area-based} and \emph{feature-based} categories \cite{zitova2003image} \cite{brown1992survey}.  Feature-based approaches rely on detection of \emph{keypoints} (points, edges, regions, corners, local gradients, etc.) which are then corresponded to estimate motion transformation parameters.  In astronomy, these methods tend to be focused on matching starfields \cite{beroiz2020astroalign} \cite{lang2010astrometry} or geophysical imagery \cite{ma2010fully} and the keypoint detection algorithms generally perform poorly for astronomical imagery which is smoothly varying or very low SNR.  In these situations, area-based methods which take a statistical approach are more appropriate.


There have been several publications which propose a maximum likelihood (ML) approach to registration, starting with Bradley \cite{bradley1973equivalence}, who showed an equivalence between ML and least squares estimates along with Kumar \cite{kumar1992correlation} specifically for registration under additive white Gaussian noise (AWGN).  Later, Mort \& Srinath \cite{mort1988maximum} developed an ML subpixel registration algorithm for pairs of Nyquist-sampled images.  Guillaume, et al. \cite{guillaume1998maximum} and Gratadour et al. \cite{gratadour2005sub} developed a method for a sequence of low SNR astronomical image frames under Poisson and Gaussian noise.


Our algorithm \footnote{https://github.com/evidlo/multiml} is most closely related to to the method devised by Gratadour et al. \cite{gratadour2005sub}, who applied their algorithm to images from an infrared galactic source taken from a ground based telescope.  They derive a cost function for an image sequence with unconstrained motion, then used an iterative conjugate gradient minimization requiring evaluation of the cost gradient along with hyperparameter tuning.

We have discovered that when this cost function is constrained to constant interframe motion, the global optimum can be found without the use of an iterative minimizer, requiring only image addition, multiplication, downscaling, and FFT operations which are easily implementable on embedded platforms like FPGAs or are already available as off-the-shelf IP cores \cite{xilinx}, making on-board registration and fusion much simpler.


\begin{figure}[htb]
\begin{minipage}[b]{.48\linewidth}
  \centering
  \centerline{\includegraphics[width=4.0cm]{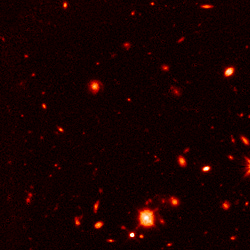}}
  \centerline{Noiseless frame}\medskip
\end{minipage}
\hfill
\begin{minipage}[b]{0.48\linewidth}
  \centering
  \centerline{\includegraphics[width=4.0cm]{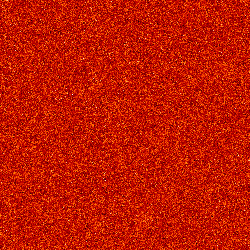}}
  \centerline{Noisy frame. -25dB SNR}\medskip
\end{minipage}
  \caption{Simulated spacecraft field of view.  False color.}
\label{fig:scene}
\end{figure}

Our algorithm was conceived for the upcoming VISORS mission, a technology demonstration of a diffractive optical element known as a photon sieve designed to study the solar corona at high resolution \cite{gundamraj2021preliminary} \cite{koenig2021formation}. VISORS features two freely-flying cubesats where the apparent motion of the scene is approximately constant translation during the 10 second science capture window.  This stronger assumption about the motion allows us to register image frames at lower SNRs than more general methods, even when visible features are significantly below the noise floor, as in Fig. \ref{fig:scene}.  Additionally, registering and fusing images sequences on-board rather than on the ground can significantly increase the science return with the limited data downlink budget available on VISORS.


In the next few sections, we provide an observation model describing motion of the frames and noise, a description of the algorithm and implementation for fast computation, a proof of optimality under the described motion and noise model, and experimental results comparing registration error of the algorithm with other area-based astronomical registration methods.

\section{Observation Model and Algorithm}
\label{sec:algorithm}

Let $\bm{y}_1, ..., \bm{y}_K \in \mathbb{R}^{N \times N}$ be an ordered sequence of $K$ noisy observed frames captured at a constant frame rate with constant drift between frames of $\bm{c}=[c_1, c_2]^T$ pixels.  Each frame has an offset of $k\bm{c}$ relative to some unknown ground truth.

For each $\bm{y}_k$, we have

\begin{equation}
\bm{y}_k = T_{k\bm{c}}(\bm{\mu}) + \bm{n}_k
\label{eq:model}
\end{equation}

where $T_{k\bm{c}}$ is a translation operator by vector $k\bm{c}$ pixels, $\bm{\mu} \in \mathbb{R}^{N \times N}$ is a Nyquist sampled version of the ground truth scene (also known as the reference image), and $\bm{n}_k \in \mathbb{R}^{N \times N}$ is measurement noise.  We have assumed $T_{k\bm{c}}$ to be a circular translation for ease of derivation, which holds approximately true for small motion vector $\bm{c}$ relative to the size of a frame and has been addressed in other literature for larger values of $\bm{c}$ by windowing the images in a preprocessing step \cite{cain2001projection}.

Note that if blurring induced by motion and the imaging system is spatially and temporally invariant, this constant PSF may be incorporated into $\bm{\mu}$ and accounted for after registration.

The maximum likelihood solution is given by

\begin{equation}
\hat{\bm{c}} = \arg \max_{\bm{c}} \sum_{m=1}^{K-1}\sum_{k=1}^{K-m} (\bm{y}_k \lstar \bm{y}_{k+m})[m\bm{c}]
\label{eq:algorithm_orig}
\end{equation}

which consists of a series of correlations denoted by $\star$, two summations, and a downsampling by $m$.  Taking $\arg \max$ over the resultant surface yields the maximum likelihood estimate of the motion, $\hat{\bm{c}}$.  The number of multiplications required is on the order of $O(N^4K^2).$

\begin{figure}[htb]
  \begin{minipage}[b]{1\linewidth}
    \centering
    \centerline{\includegraphics[width=8.5cm]{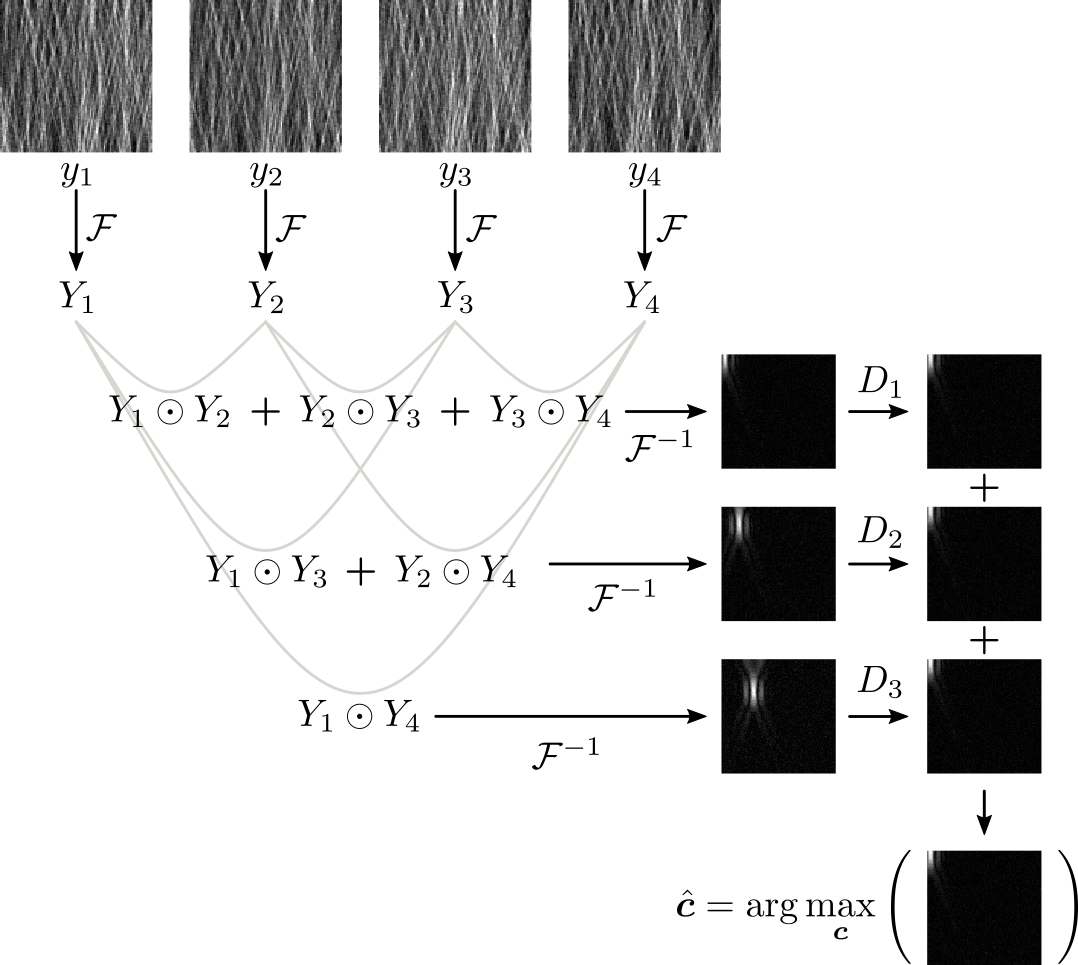}}
  \end{minipage}
  \caption{Graphical diagram of algorithm given in Equation \ref{eq:algorithm} for a sequence of 4 frames}
  \label{fig:algorithm}
\end{figure}

The algorithm can be accelerated by performing correlations in the frequency domain and precomputing the Fourier transforms of the images with FFTs:

\begin{equation}
\hat{\bm{c}} = \arg \max_{\bm{c}} \sum_{m=1}^{K-1} D_m \left[
\mathcal{F}^{-1} \left( \sum_{k=1}^{K-m} \bm{Y}_k \odot \bm{Y}_{k+m} \right)
\right][\bm{c}]
\label{eq:algorithm}
\end{equation}

where $D_m$ is a downsample operator by $m$ pixels (with zero padding to maintain shape), $\mathcal{F}^{-1}$ is the inverse FFT, $\bm{Y}_k$ is the Fourier transform of $\bm{y}_k$, and $\odot$ is the elementwise product operator.  The new complexity is $O(KN^2\log N + K^2N^2)$.
The algorithm can be broken into four steps, shown below and illustrated graphically in Fig. \ref{fig:algorithm}.

\begin{enumerate}
  \item Precompute image Fourier transforms:
    $$\bm{Y}_k=\mathcal{F}(\bm{y}_k) \text{ for } k=1,...,K$$
  \item Compute image correlations and sum into groups by degree of separation $m$:
    $$\bm{S}_m = \mathcal{F}^{-1} \left( \sum_{k=1}^{K-m} \bm{Y}_k \odot \bm{Y}_{k+m} \right) \text{ for } m=1, ..., K-1$$
  \item Downsample correlation groups by degree of separation and sum:
    $$\sum_{m=1}^{K-1} D_m \left[ \bm{S}_m \right]$$
  \item Take the argmax of the resultant surface to find the estimate $\hat{\bm{c}}$
    $$\hat{\bm{c}} = \arg \max_{\bm{c}} \sum_{m=1}^{K-1} \left(D_m \left[ \bm{S}_m \right]\right)[\bm{c}]$$
\end{enumerate}


\section{Proof of ML Optimality}
\label{sec:optimality}

The proof of optimality is presented in 3 parts:

\begin{enumerate}
\item Derive the expression for likelihood maximization over $\bm{c}$ and $\bm{\mu}$
\item Derive the most likely value for $\bm{\mu}$ as a function of $\bm{c}$
\item Show that the log-likelihood solution consists of a sum of downsampled cross correlations
\end{enumerate}

Without loss of generality, we assume the observed frames to be one dimensional vectors of length $N$ and that drift $c$ is a scalar.

Given the observation model

\begin{equation*}
y_k = T_{kc}(\mu) + n_k
\end{equation*}

assume $n_k \sim \mathcal{N}(0, \sigma^2)$ is additive white Gaussian noise with variance $\sigma^2$.  We note that the proof may be extended to non-stationary noise, similar to \cite{gratadour2005sub}, but omit it here for brevity.


The log-likelihood of having a particular $c$ and $\mu$ given observation sequence $y_1, ..., y_K$ is

\begin{align}
  \hat{c}, \hat{\mu} &= \arg \max_{c, \mu} \ln \mathcal{L}(c, \mu \lpipe y_1, ..., y_K) \nonumber \\
  &=\arg \max_{c, \mu} \ln \prod_{k=1}^K \prod_{n=1}^N \tfrac{1}{\sigma^2 \sqrt{2 \pi}} \text{exp} \left[- \frac{(y_{k,n} - T_{kc}(\mu)_n)^2}{2\sigma^2}\right] \nonumber \\
  &=\arg \min_{c, \mu} \underbrace{\sum_{k=1}^K \sum_{n=1}^N (y_{k,n} - T_{kc}(\mu)_n)^2}_{\text{cost}(c, \mu)} \label{eq:min}
\end{align}

We will take the derivative of the expression denoted cost($c$, $\mu$) in Equation \ref{eq:min} with respect to the $j$th element of $\mu$ to eliminate minimization over $\mu$.


\begin{align}
  &\frac{d}{d\mu_j}\left[\text{cost}(c, \mu)\right]
  =\frac{d}{d\mu_j}\left[\sum_{k=1}^K \sum_{n=1}^N (y_{k,n} - T_{kc}(\mu)_n)^2\right] = 0 \nonumber  \\
  &\Longrightarrow \hat{\mu}_j = \sum_{k=1}^K T_{-kc}(y_k)_j
  \Longrightarrow \hat{\mu} = \sum_{k=1}^K T_{-kc}(y_k) \label{eq:mu}
\end{align}

This is simply the sum of the motion-corrected noisy frames.  Plugging this into the cost function, we obtain an expression that depends only on $c$.  Expanding and eliminating constant terms, we get

\begin{align*}
  &\text{cost}(c, \hat{\mu}) = \text{cost}(c) \\
  &=\sum_{k=1}^K \sum_{n=1}^N \left(y_{k,n} - T_{kc}\left(\sum_{l=1}^K T_{-lc}(y_l)\right)_n\right)^2 \\
  &=\sum_{k=1}^K \sum_{n=1}^N \cancel{y_{k,n}^2} -2y_{k,n}\sum_{l=1}^K T_{(k-l)c}(y_l)_n + \left[\sum_{l=1}^K T_{(k-l)c}(y_l)_n\right]^2 \\
  &=\sum_{\substack{k=1 \\ l=1}}^K \sum_{n=1}^N -2y_{k,n} T_{(k-l)c}(y_l)_n +
  \sum_{k=1}^K \sum_{n=1}^N \left[\sum_{l=1}^K T_{(k-l)c}(y_l)_n\right]^2 \\
\end{align*}

Examining the first term in the cost function, we see it is simply a downsampled correlation between $y_k$ and $y_l$.

$$\sum_{\substack{k=1 \\ l=1}}^K \sum_{n=1}^N -2y_{k,n} T_{(k-l)c}(y_l)_n = -2\sum_{\substack{k=1 \\ l=1}}^K (y_k \lstar y_l) [(k-l)c]$$

Similarly, the second term can be manipulated into a cross correlation.

\begin{align*}
  &\sum_{k=1}^K \sum_{n=1}^N \left[\sum_{l=1}^K T_{(k-l)c}(y_l)_n\right]^2 \\
  &= \sum_{k=1}^K \sum_{n=1}^N \left[\sum_{l=1}^K \left( T_{(k-l)c}(y_l)_n \sum_{m=1}^K T_{(k-m)c}(y_m)_n \right)\right] \\
  &= \sum_{k=1}^K \sum_{n=1}^N \sum_{l=1}^K \sum_{m=1}^K T_{(k-l)c}(y_l)_n T_{(k-m)c}(y_m)_n \\
  &= K \sum_{k=1}^K \sum_{l=1}^K (y_k \lstar y_l)[(k-l)c]
\end{align*}

Combining the two terms, we arrive at

\begin{align*}
  \hat{c} &= \arg \min_c \text{cost}(c) \\
  &= \arg \max_c (K - 2) \sum_{k=1}^K \sum_{l=1}^K (y_k \lstar y_l)[(k - l)c] \\
  &= \arg \max_c \sum_{m=1}^{K-1} \sum_{k=1}^{K-m} (y_k \lstar y_{k+m})[mc]
\end{align*}

which is identical to Equation \ref{eq:algorithm_orig} given in the previous section.

\section{Experimental Results}
\label{sec:results}

\begin{figure}[htb]
\begin{minipage}[b]{.48\linewidth}
  \centering
  \centerline{\includegraphics[width=4.0cm]{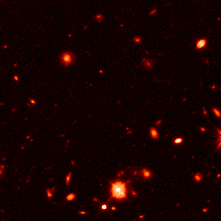}}
  \centerline{(a) Ground truth}\medskip
\end{minipage}
\hfill
\begin{minipage}[b]{0.48\linewidth}
  \centering
  \centerline{\includegraphics[width=4.0cm]{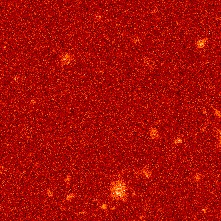}}
  \centerline{(b) Reconstruction}\medskip
\end{minipage}
  \caption{Ground truth and non-regularized reconstruction result for -25dB SNR AWGN and $K=30$ frames.}
\label{fig:recon}
\end{figure}

To evaluate performance of the algorithm, we generated a series of simulated $250\times250$ pixel noisy frames from Hubble deep field images (shown in Fig. \ref{fig:recon}a) according to the motion and noise model given in Equation \ref{eq:model}.  We estimated interframe motion using our algorithm in Equation \ref{eq:algorithm} and then coadded the corrected frames as in Equation \ref{eq:mu} to obtain the reconstruction in Fig. \ref{fig:recon}b.

\begin{figure}[htb]
  \begin{minipage}[b]{1\linewidth}
    \centering
    \centerline{\includegraphics[width=8.5cm]{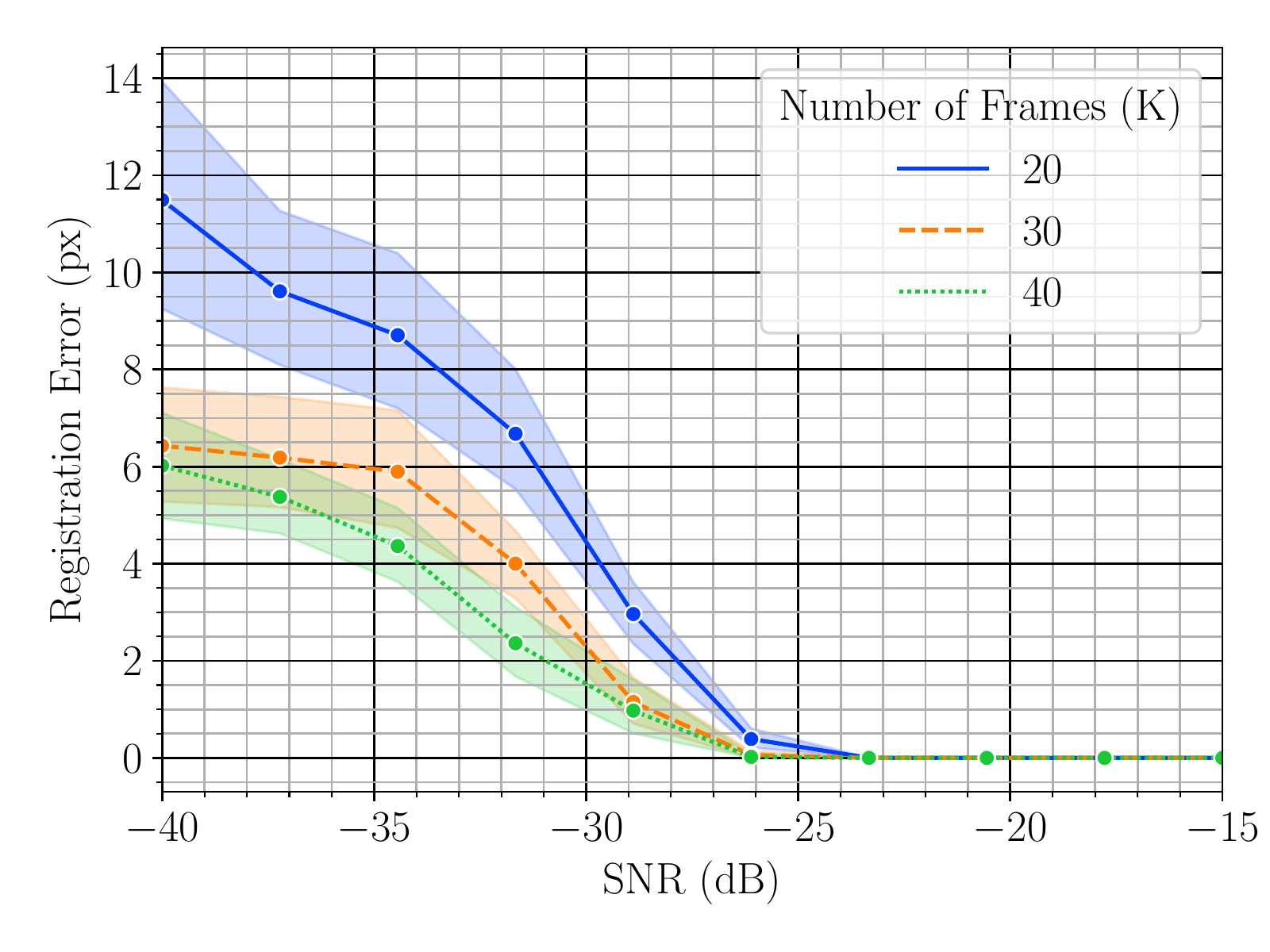}}
  \end{minipage}
  \caption{Registration absolute error for various AWGN SNRs and number of frames $K$ with a randomly chosen constant motion vector. 50 trials. 1 std. dev. error bands.}
  \label{fig:db_sweep}
\end{figure}

We repeated this experiment with various noise levels and number of frames for 50 trials each with a random motion vector and noise.  As shown in Fig. \ref{fig:db_sweep} and corroborated in \cite{gratadour2005sub}, the registration error decreases with increasing number of frames $K$ and is able to obtain less than 1 pixel of mean absolute registration error at -25dB measurement noise for $K=20$ frames, and the same at -30dB for $K=40$.

Next, we compared registration performance on astronomical data against another area-based registration method provided by Ginsburg et al. \cite{ginsburg2013bolocam}, originally used for registering images of cosmic dust in infrared and also against a pairwise version of the algorithm given in \cite{guizar2008efficient}.  Since these algorithms have a non-constant motion model, we project their motion estimates to the nearest constant, rigid motion estimate for a fairer comparison.  Again, the experiment was repeated for 50 trials with a randomized motion vector and noise realization.

\begin{figure}[htb]
  \begin{minipage}[b]{1\linewidth}
    \centering
    \centerline{\includegraphics[width=8.5cm]{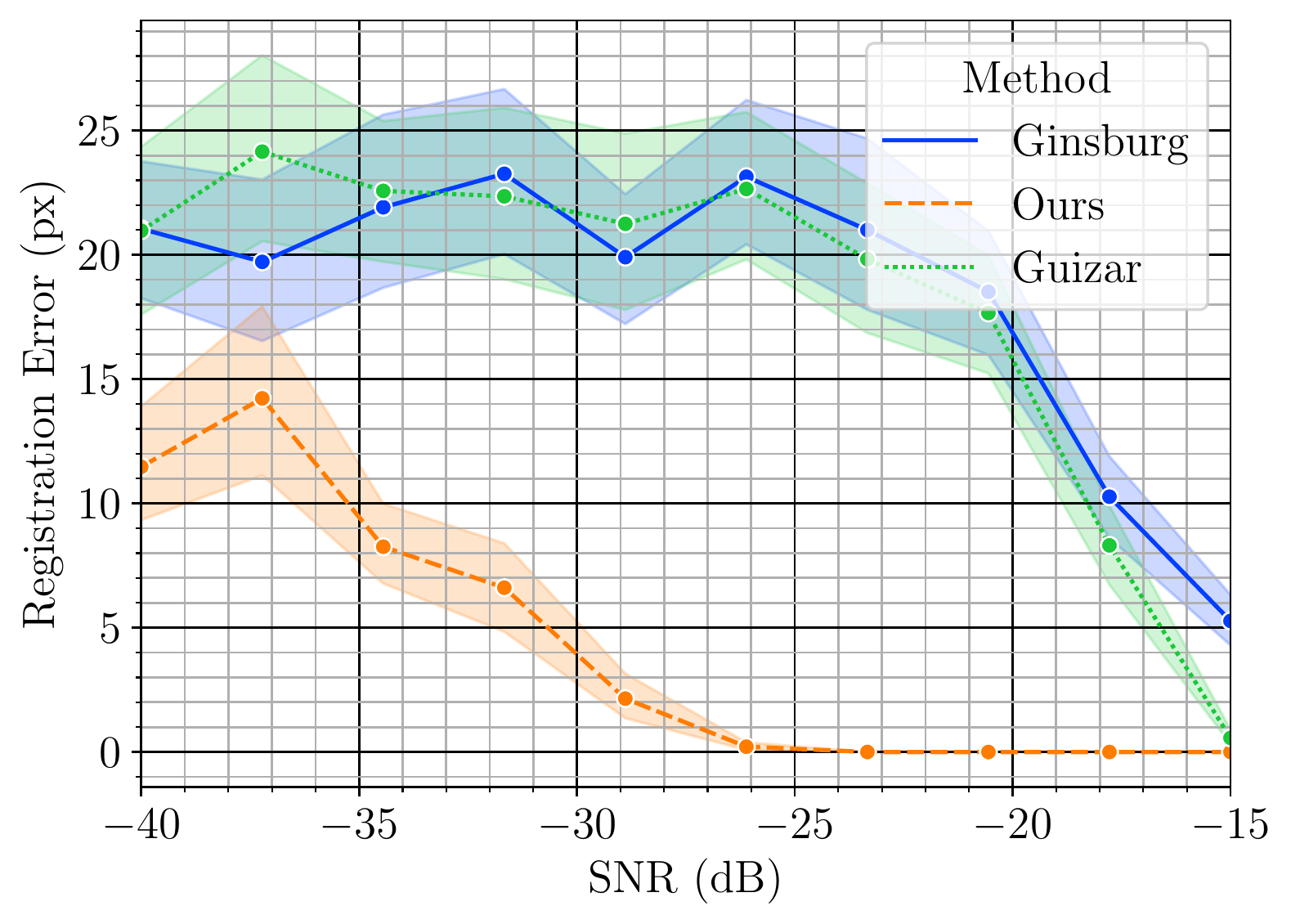}}
  \end{minipage}
  \caption{Comparison of registration error of our method vs the one presented in \cite{ginsburg2013bolocam} for various noise levels, $K=20$. 50 trials.  1 std. dev. error bands.}
  \label{fig:method_compare}
\end{figure}

Fig. \ref{fig:method_compare} shows that our algorithm is able to successfully register images at over 10dB lower SNR.

\section{Summary and Conclusion}
\label{sec:future}

In this manuscript, we presented a multiframe registration algorithm for constant rigid motion.  We showed that the algorithm can be realized without iterative methods or parameter tuning, and that the algorithm is optimal in the maximum likelihood sense.  We characterized the algorithm for various noise levels and image sequence lengths.

The algorithm may be directly extended to the subpixel domain by applying the techniques described in \cite{guizar2008efficient}.  It is also possible to handle constant scaling and rotation using the Log-Polar transform as described in \cite{reddy1996fft}, but this requires interpolation that may be expensive on embedded platforms.

The algorithm is useful in settings where motion is constant and rigid, images are low SNR, or a straightforward implementation on an embedded system is needed.

\clearpage
\vfill\pagebreak

\bibliographystyle{IEEEbib}
\bibliography{refs}

\end{document}